\documentclass[twocolumn,showpacs,preprintnumbers,amsmath,amssymb]{revtex4}
\usepackage{graphicx}% Include figure files
\usepackage{dcolumn}% Align table columns on decimal point
\usepackage{bm}% bold math
\usepackage{epsf,epsfig,latexsym}

\begin{document}

\preprint{APS/123-QED}

\title{Caloric Curves and Critical Behavior in Nuclei}

\author{J. B. Natowitz}
\author{R. Wada}
\author{K. Hagel}
\author{T. Keutgen}
\author{M. Murray}
\author{A. Makeev}
\author{L. Qin}
\author{P. Smith}
\author{C. Hamilton}

\affiliation{Cyclotron Institute, Texas A\&M University,\\
  College Station,  Texas, 77845}%

\date{\today}% It is always \today, today           %  but any date may be explicitly specified

\begin{abstract}
   Data from a number of different experimental measurements have been used to
   construct caloric curves for five different regions of nuclear mass. These
   curves are qualitatively similar and exhibit plateaus at the higher excitation
   energies. The limiting temperatures represented by the plateaus decrease with
   increasing nuclear mass and are in very good agreement with results of recent
   calculations employing either a chiral symmetry model or the Gogny interaction.
   This agreement strongly favors a soft equation of state. Evidence is presented
   which suggests that critical excitation energies and critical temperatures might
   be determined from caloric curve measurements when the mass variations inherent
   in such measurements are taken into account.
\end{abstract}

\maketitle

 \section{Introduction}

   Measurements of the nuclear specific heat have long been considered to be a
   technique that should provide important information on the properties of excited
   nuclei and the postulated liquid-gas phase transition
   \cite{Suraud89,Bonche86,Gross93,Bondorf85,Friedman88}.
   Over slightly more
   than a decade a number
   of measurements have been motivated by this expectation
\cite{Hagel,Wada89,Cussol,Chulick,Gonin90,Pochadzalla,Odeh99,Hauger00,%
 Wada97,Kwiatkowsk98, Morley96,Cibor00,Hagel00,Ruangma}.
   However, given
   the significant variation in the systems studied, in the collision dynamics
   involved, in the experimental and analysis techniques employed, in the
   theoretical  descriptions proposed and even in the way the results are reported,
   a coherent picture of the information content in caloric curves has been
   difficult to obtain. Indeed, two recent reviews of caloric curve measurements
\cite{Borderie,DasGupta01}
   have reached rather pessimistic conclusions on the utility of such
   measurements. We present evidence that, in fact, the existing body of data
   provides a rather consistent picture when the mass dependence of the caloric
   curve measurements is taken into account. Further, comparison with the results
   of a recently reported Fisher Droplet Model analysis establishing the critical
   point in the A$\sim$160 region
\cite{Elliott01}
   indicates that the available caloric curve data
   provide  direct measures of both the critical energy and the critical
   temperature for the phase change into a {\it  non-monomeric} gaseous phase over a wide
   range of nuclear masses.

\section{Analysis and Results}

\subsection{Selection of Data}

   We present in this paper an analysis of the combined results of temperatures and
   excitation energies of
   references \cite{Hagel,Wada89,Cussol,Chulick,Gonin90,Pochadzalla,Odeh99,Hauger00,
 Wada97,Kwiatkowsk98, Morley96,Cibor00,Hagel00,Ruangma}.
 We have selected these results because
   in each of the cases considered the authors have attempted a simultaneous
   derivation of the {\it initial} temperatures of de-exciting nuclei of  reasonably well
   characterized nuclear masses and excitation energies. Except at the lowest
   projectile energies, a de-convolution of product energy and yield spectra is
   usually required in order that these properties may be established. Typically,
   phenomenological or theoretical corrections must be applied to measured values
   to obtain the desired initial values. This is usually necessary in the case of
   the measured {\it apparent} temperatures, whether slope temperatures
 \cite{Hagel,Wada89,Cussol,Chulick,Gonin90}
   or double
   isotope yield ratio temperatures
\cite{Pochadzalla,Odeh99,Hauger00,%
 Wada97,Kwiatkowsk98, Morley96,Cibor00,Hagel00,Ruangma,Albergo85},
   to account for de-excitation
   cascades and secondary particle contributions.

   To obtain the initial thermal excitation energies, corrections for unobserved
   ejectiles, i.e. neutrons, gamma rays (small and sometimes neglected) and
   undetected charged species
   are often needed.  In one case, statistical model calculations have been
   employed to ``backtrace" the excitation energy \cite{Odeh99}.
   Where corrections
   to the raw observed values are required, they have been applied in the
   referenced works or sufficient information has been given to allow such
   corrections to be made for the present analysis. In Table 1 the experimental
   investigations which are the sources of the data included in our analysis are
   listed together with an indication of the techniques used to extract and to
   correct the excitation energies and temperatures. Much more detail on the
   methods employed to analyze the experiments and to make the corrections required
   to determine the excitation energies and temperatures of the primary hot
   composite nuclei is presented in the references
 \cite{Hagel,Wada89,Cussol,Chulick,Gonin90,Pochadzalla,Odeh99,Hauger00,%
 Wada97,Kwiatkowsk98, Morley96,Cibor00,Hagel00,Ruangma} . Below we make only
   a few additional comments on some of these papers to better explain our use of
   the available information found there. We emphasize that the goal in each case
   is to determine the excitation energy and temperature of the primary composite
   system that remains after early non-equilibrium emission processes subside.

   1. The Aladin Collaboration has determined temperatures by multiplying observed
   double isotope, $T_{HeLi}$, temperatures by a factor of 1.2
    \cite{Pochadzalla}.
    This factor,
   intended to correct the observed data for effects of secondary emission data,
   was determined from Quantum Statistical Model calculations. While the initially
   reported caloric curve for this system relied on calorimetric techniques for
   energy determination, more recently, in reference \cite{Odeh99},
   a back-tracing technique
   was employed to determine the thermal excitation energy after the early
   non-equilibrium emission phase of the reaction. This is done by requiring that
   the mass distributions and other observables, calculated using the SMM model,
   agree with the measured distributions. We utilize this more recent thermal
   energy determination and the caloric curve presented in Figure 61 of reference
   \cite{Odeh99}.

   2. In reference \cite{Hauger00}, Hauger {\it et al.} present results for their analysis of the data
   for 1GeV/nucleon  $^{84}$Kr, $^{139}$La and $^{197}$Au beams on $^{12}C$ targets. Initial Excitation
   energies and masses were obtained by subtracting the energy and mass removed in
   the pre-equilibrium stage of the reaction (see Figures 12 and 18 of Ref
     \cite{Hauger00}.)
   {\it Initial} temperatures were derived from the SMM model calculations which employed
   the experimental initial excitation energies and masses and were found to be in
   good agreement with the final observed distributions. Thus it is the ``hot
   caloric curves" presented in Figure 20 of reference 13 which are used in
   this paper.

   3.  For the experiments of the ISiS collaboration
   \cite{Kwiatkowsk98, Morley96, Ruangma}
   initial
   excitation energies and masses are also obtained by subtracting the energy and
   mass removed in the pre-equilibrium stage of the reaction. Raw temperatures are
   determined with a relatively high selection of ejectile energy range in the
   remaining spectrum. Some sensitivity to the selected range is observed. Since
   the pre-equilibrium component has been removed this sensitivity
   appears to be indicative of
   cooling in the investigated systems.  In references
    \cite{Wada97,Kwiatkowsk98}
   which report
   measurements for $^{3}$He projectiles with $^{197}$Au and Ag, temperatures are not
   corrected for this effect. However, comparisons of the results of SMM
   calculations for the $^{3}$He +$^{197}$Au system, with and without energy cuts equivalent
   to those used in the experiments, are presented. The comparison allows
   estimation of the factors required to correct the apparent temperatures. We have
   corrected the Ag data by assuming the same correction factors apply.
   In reference \cite{Ruangma} the temperatures are corrected for secondary decay using
   parameters suggested by Tsang {\it et. al.}  \cite{Tsang}. These corrections are relatively
   small, typically less than 10\%.

   4. The temperatures of references
    \cite{Cibor00, Hagel00}
    are established for identical
   velocity species in a coalescence type of analysis. No temperature correction is
   applied as the techniques employed should be effective at discriminating against
   secondary decay contributions.  Excitation energies are determined
   calorimetrically.

   Only some of the experiments summarized in Table 1 have measured neutron
   multiplicities   \cite{Gonin90,Pochadzalla,Odeh99,Wada97,Cibor00,Hagel00}
and only two  have measured neutron
   spectra
   \cite{Gonin90,Odeh99}.
   The other experiments employing calorimetric techniques determine the
   neutron emission contribution to the thermal energy using phenomenological
   corrections based upon related observations and/or statistical calculations. At
   low excitations where neutron emission dominates this can lead to larger
   uncertainties \cite{Lefort}. It may also lead to systematically larger uncertainties at
   the higher limits of the apparent excitation energy spectrum where fluctuations
   may be significant \cite{Goldenbaum}. In general both excitation energies and temperatures
   appear to be subject to systematic uncertainties of $\approx 10$\% in the various
   experiments \cite{Goldenbaum}.

   Additional information on caloric curves has been reported by the INDRA
   collaboration \cite{Ma97,Peter98a,Peter98b}.
   It is our understanding that the excitation energy
   determinations and temperatures for those experiments are currently under review
    \cite{Dore} and therefore we have not included them in this work.

\subsection{Correlation of Temperature, Excitation Energy and Mass}

   In Fig.~\ref{fg:SumCalor} we plot the correlated values of the temperatures and excitation
   energies per nucleon which have been determined in the experiments represented
   in Table 1. For comparison we  plot curves corresponding to Fermi-gas model
   predictions with inverse level density parameters K=8 and K=13. Also included
   for a further reference is a  ``total vaporization" line representing
   the (purely hypothetical) two-stage scenario of separation into constituent
   nucleons, at a cost of 8 MeV/nucleon followed by thermal heating. Although there
   is a significant increase in divergence of the results at higher excitation
   energies, the combined data still exhibit the qualitative features observed
   previously in many of the individual experiments,  i.e. an apparent Fermi
   gas-like rise at excitation energies per nucleon below 3-4 MeV/nucleon, a very
   slow rate of temperature increase at higher energies ($\sim4-9$MeV/nucleon) and
   some indication of an increase again at higher energy, $\sim9$ MeV/nucleon.

   In some of the experiments considered the more central collisions were selected
   by an appropriate experimental filter and the excitation energy was varied by
   changing the projectile energy
   \cite{Chulick,Gonin90,Cibor00,Hagel00}.
   In such cases the masses of the
   excited systems
   ( following pre-equilibrium emission) are reasonably well constrained. In other
   experiments a single projectile energy was used and a range of impact parameters
   from peripheral to central was investigated
    \cite{Hagel,Wada89,Cussol,Pochadzalla,Odeh99,Hauger00,%
    Wada97,Kwiatkowsk98,Morley96}.
    In those cases the
   initial masses, and excitation energies of the excited systems studied may vary
   significantly with impact parameter. That the results included in Fig.~\ref{fg:SumCalor}
   include experiments which span a broad range of mass is illustrated in
   Figs.~\ref{fg:ExvMass}
   and ~\ref{fg:TvMass} which present plots of the primary excitation energies and temperatures as
   a function of the derived values of the primary mass. The wide mass variation
   inherent in many of the individual experiments is clear as is the fact that
   different experiments may sample the same mass range at significantly different
   excitation energies. We have previously suggested that mass variation is an
   important factor which should be considered in any interpretation of the caloric
   curve \cite{Natowitz95a}.

\subsection{Caloric Curves for Restricted Mass regions}

\begin{figure}
\epsfig{file=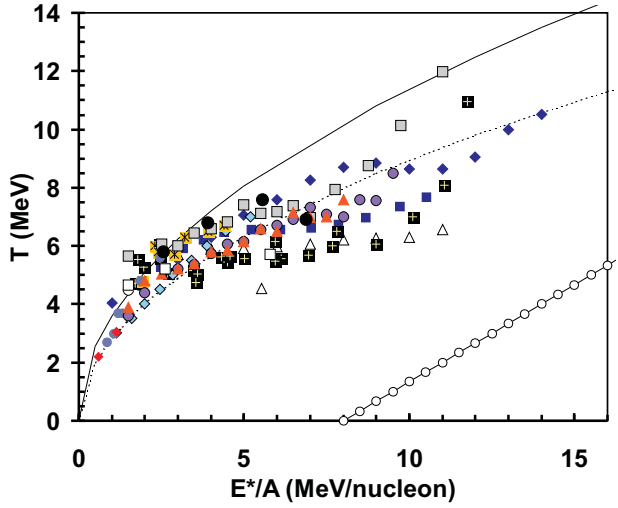,width=9.2cm,angle=0}
\caption{\label{fg:SumCalor}
   Caloric curve data from references
   \cite{Hagel,Wada89,Cussol,Chulick,Gonin90,Pochadzalla,Odeh99,Hauger00,%
 Wada97,Kwiatkowsk98, Morley96,Cibor00,Hagel00,Ruangma}
  Measurements of  temperature vs
                        excitation energy per nucleon are represented by symbols.
Reference  \cite{Hagel}-dark gray filled circles;
Reference  \cite{Wada89} -filled gray squares with  black stars;
Reference  \cite{Cussol} - filled light-gray diamonds;
Reference  \cite{Chulick} - filled dark-gray diamonds;
Reference  \cite{Gonin90} - open circles;
References \cite{Pochadzalla,Odeh99} - black filled squares with white plus signs;
Reference  \cite{Hauger00} - black diamonds ($^{84}$Kr),
   black squares ($^{134}$La) and open triangles($^{197}$Au);
Reference  \cite{Wada97} - Black triangles;
References \cite{Kwiatkowsk98, Morley96} - light-gray filled circles (Ag);
   light-gray filled triangles ($^{197}$Au);
References \cite{Cibor00,Hagel00} - black filled circles;
Reference \cite{Ruangma} - Light- gray filled squares.
Fermi-gas model lines for K=8 (dashed line)
and K=13 (solid line) and a ``total vaporization"
   line (see text) - connected small open circles are shown for comparison. }
 \end{figure}

\begin{figure}
%%\center{\epsfig{file=fig2.eps,width=17.0cm,angle=0}}
\epsfig{file=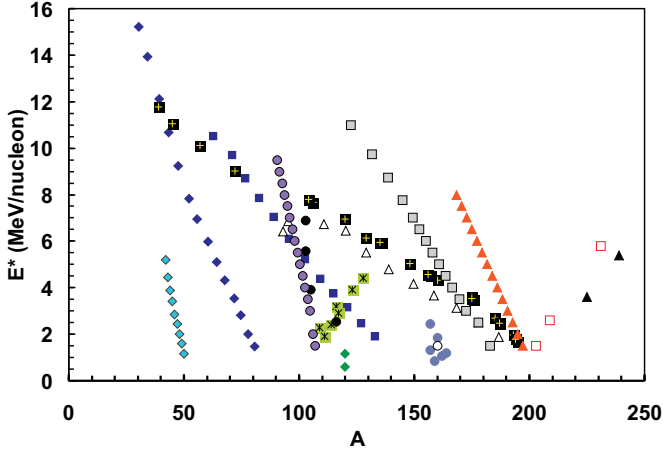,width=9.4cm,angle=0}
\caption{\label{fg:ExvMass}  Excitation Energy per nucleon as a function of A.
Reference  \cite{Hagel}-dark gray filled circles;
Reference  \cite{Wada89} -filled gray squares with  black stars;
Reference  \cite{Cussol} - filled light-gray diamonds;
Reference  \cite{Chulick} - filled dark-gray diamonds;
Reference  \cite{Gonin90} - open circles;
References \cite{Pochadzalla,Odeh99} - black filled squares with white plus signs;
Reference  \cite{Hauger00} - black diamonds ($^{84}$Kr),
   black squares ($^{134}$La) and open triangles($^{197}$Au);
Reference  \cite{Wada97} - Black triangles;
References \cite{Kwiatkowsk98, Morley96} - light-gray filled circles (Ag);
   light-gray filled triangles ($^{197}$Au);
References \cite{Cibor00,Hagel00} - black filled circles;
Reference \cite{Ruangma} - Light- gray filled squares.
}
\end{figure}

\begin{figure}
%\center{\epsfig{file=fig3.eps,width=17.0cm,angle=0}}
\epsfig{file=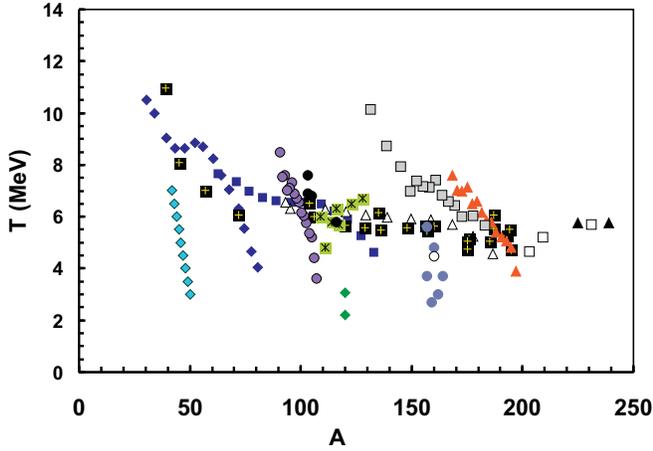,width=9.4cm,angle=0}
\caption{
\label{fg:TvMass}
Temperature as a function of A, the mass number of the primary
   de-exciting
                     nucleus.
Reference  \cite{Hagel}-dark gray filled circles;
Reference  \cite{Wada89} -filled gray squares with  black stars;
Reference  \cite{Cussol} - filled light-gray diamonds;
Reference  \cite{Chulick} - filled dark-gray diamonds;
Reference  \cite{Gonin90} - open circles;
References \cite{Pochadzalla,Odeh99} - black filled squares with white plus signs;
Reference  \cite{Hauger00} - black diamonds ($^{84}$Kr),
   black squares ($^{134}$La) and open triangles($^{197}$Au);
Reference  \cite{Wada97} - Black triangles;
References \cite{Kwiatkowsk98, Morley96} - light-gray filled circles (Ag);
   light-gray filled triangles ($^{197}$Au);
References \cite{Cibor00,Hagel00} - black filled circles;
Reference \cite{Ruangma} - Light- gray filled squares.
}
\end{figure}

\begin{figure}%[h]
\epsfig{file=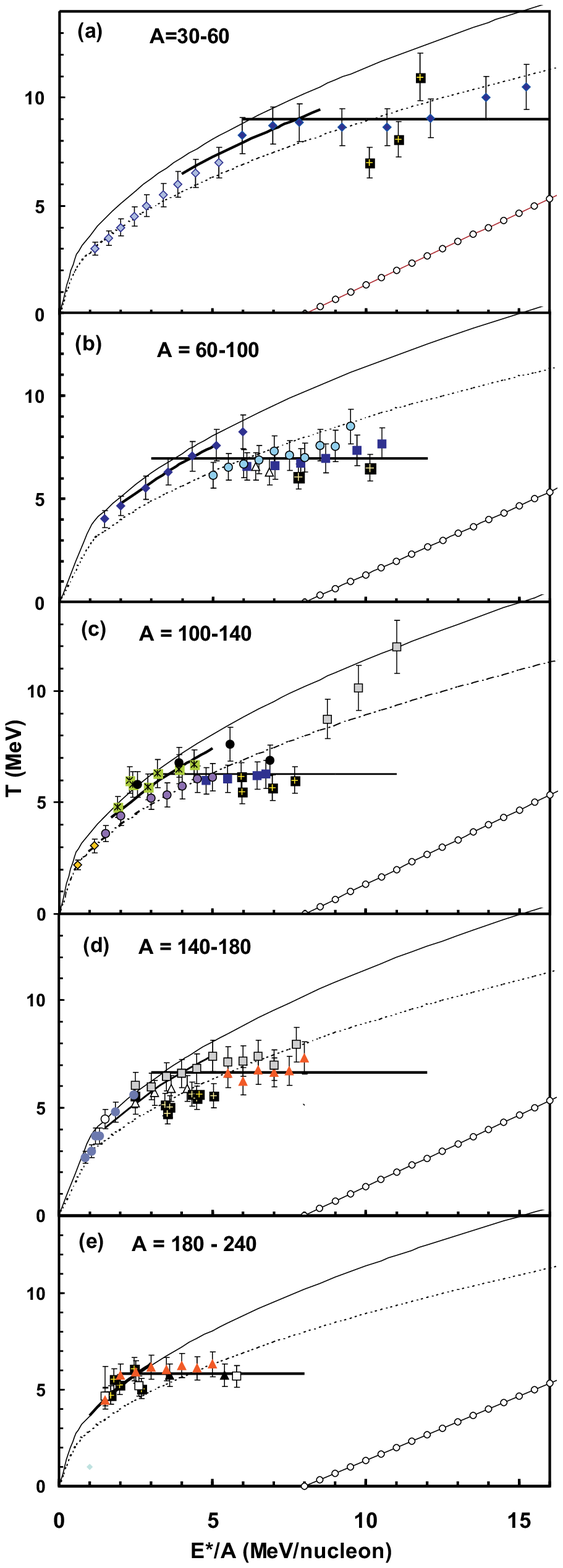,width=9.4cm,angle=0}
\caption{\label{fg:CalorvMass}
 Caloric curves for five selected regions of mass.
 Reference  \cite{Hagel}-dark gray filled circles;
Reference  \cite{Wada89} -filled gray squares with  black stars;
Reference  \cite{Cussol} - filled light-gray diamonds;
Reference  \cite{Chulick} - filled dark-gray diamonds;
Reference  \cite{Gonin90} - open circles;
References \cite{Pochadzalla,Odeh99} - black filled squares with white plus signs;
Reference  \cite{Hauger00} - black diamonds ($^{84}$Kr),
   black squares ($^{134}$La) and open triangles($^{197}$Au);
Reference  \cite{Wada97} - Black triangles;
References \cite{Kwiatkowsk98, Morley96} - light-gray filled circles (Ag);
   light-gray filled triangles ($^{197}$Au);
References \cite{Cibor00,Hagel00} - black filled circles;
Reference \cite{Ruangma} - Light- gray filled squares.
}
\end{figure}

   In order to explore the extent to which the variations in reported caloric
   curves seen in Fig.~\ref{fg:SumCalor} are affected by mass variation we have constructed
   composite caloric curves for five different mass regions by combining the
   appropriate data from each of the systems of Table~\ref{tb:DataRef}. In Fig.~\ref{fg:CalorvMass} (a)-(e) we
   present the resultant curves for the mass number regions 30-60,
   60-100,100-140,140-180 and 180-240. For comparison each sub-figure also includes
   the calculated Fermi gas curves  for K=8 and K=13,  as well as the ``total
   vaporization" line presented in Fig.~\ref{fg:SumCalor}.

   Viewed in this way the resultant curves are qualitatively similar, in general
   rising at low energies, trending  towards the K=13 line, and then leveling into
   a plateau-like region.
   The quantitative aspects of the behavior in the lower energy region and the
   importance of the temperature dependence of the effective mass in determining
   the level density parameter have previously been extensively discussed
    \cite{Hasse,Shlomo90,Shlomo91}.
   The rise toward K=13 and flattening of the curve, representing a sharp rise in
   the heat capacity, has been discussed
   \cite{Natowitz95a,Natowitz95b}
   and compared with model  predictions. In statistical models of
   multifragmentation the break occurs at a ``cracking energy" which
   represents the onset of multiple fragment production
   \cite{Gross93,Bondorf85}.
   Within the framework
   of classical molecular dynamics calculations
    \cite{Bonasera, Strachan}
   quantum molecular dynamics
   calculations
   \cite{Sugawa,Feldmeier}
   and more microscopic treatments
   \cite{De97a,Kolomietz}
   plateaus are also
   observed.

   For the lightest mass window, A=30-60, the increase above K=8 is less
   pronounced but there is evidence of a flattening   near 8 MeV/nucleon excitation
   energy.   For the next three windows this feature appears near 4 MeV/nucleon
   excitation. For the highest window it seems to occur even lower, near 3
   MeV/nucleon. Although there is considerable spread in the data, we have
   determined the average temperatures in the plateau regions for each mass window.
   This was done by using the data at excitation energies above the points where
   the flattening appears to set in. The results of temperature measurements for
   excitation energies  $> 9$ MeV/nucleon, noted above in the discussion of
   Fig.~\ref{fg:SumCalor} as suggesting a later rise in the caloric curve, are seen in Fig.~\ref{fg:CalorvMass}
   to be dominated by the data in the lowest mass window, A=30-60, and the three
   highest excitation energy points from reference
   \cite{Ruangma},
   which fall in the A=100-140
   mass window (Fig.~\ref{fg:CalorvMass} (c) ). In the A=30-60 mass window the evidence for this
   later rise is now less compelling and the data have been used in determining the
   average. The three points at highest excitation in Fig.~\ref{fg:CalorvMass} (c) may signal a
   further rise and have not been included. For the data of reference
   \cite{Ruangma}, the
   reported uncertainties for the highest excitation energy points become larger
   than the nominal 10\% systematic value  which we have assumed for all points.
%   Also, the effect of energy cut and application of the Tsang systematics \cite{Tsang}  in
%   an excitation energy range well above that for which that systematics was
%   established  may be affecting these results.
   The  average values are shown as solid horizontal lines in
   Fig.~\ref{fg:CalorvMass} (a)-(e). We note
   that the width of the selected  mass windows still allows for some mass
   variation and individual experiments in which the mass is changing show some
   evidence of an increase in the more restricted mass windows. Nevertheless, the
   general agreement of the data with the average leads to relatively small
   standard deviations on these averages.

   In Fig.~\ref{fg:CalorvMass}
   the value of the limiting temperature reached in the plateau
   decreases with increasing mass. It has previously been suggested
    \cite{Natowitz95b,Bonasera}
    that
   the limiting temperatures which are observed in Caloric curve measurements
   represent  the ``Coulomb instability" temperatures, first calculated with a
   temperature dependent Hartree-Fock model employing a Skyrme interaction
   \cite{Bonche86, Jaqaman, Song}
   and later with other models
    \cite{De97a, Besprovany, Zhang96, Zhang99}.
   In such calculations, the
   limiting temperature, which represents the limit of equilibrium phase
   coexistence between liquid and vapor, has been designated as the point of
   Coulomb instability because, in the absence of the Coulomb forces the
   coexistence is possible up to the critical temperature of nuclear matter
   \cite{Bonche86}.
   The observed Coulomb instability temperature has been related to the
   incompressibility and critical temperature of nuclear matter \cite{Song}.
    Limiting
   temperature data for A$\sim$120 from reference  \cite{Wada89}  were found to be in best agreement
   with results of those calculations when  the SJ1 Skyrme interaction was used
     \cite{Natowitz95a,Natowitz95b}.

\begin{figure}
%\center{\epsfig{file=fig5.eps,width=9.2cm,angle=0}}
\epsfig{file=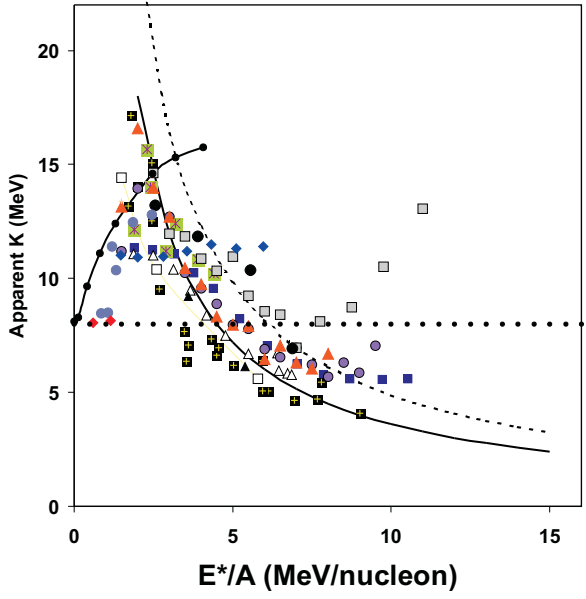,width=9.4cm,angle=0}
\caption{\label{fg:kvExA}
      Apparent Fermi gas level density parameters as a function of
   excitation
                    energy. Data are from references in Table~\ref{tb:DataRef}.  Data for nuclei
   with A$<$60 are not
                    included. (See text.) The horizontal dotted line represents a
   constant value of K= 8. The solid line with solid dots represents the theoretical
   prediction of reference \cite{Shlomo91}. Two additional lines are shown, representing the
   values of K
 corresponding to a constant T= 6 MeV ( solid line) and T = 7 MeV (dashed line).
Reference  \cite{Hagel}-dark gray filled circles;
Reference  \cite{Wada89} -filled gray squares with  black stars;
Reference  \cite{Cussol} - filled light-gray diamonds;
Reference  \cite{Chulick} - filled dark-gray diamonds;
Reference  \cite{Gonin90} - open circles;
References \cite{Pochadzalla,Odeh99} - black filled squares with white plus signs;
Reference  \cite{Hauger00} - black diamonds ($^{84}$Kr),
   black squares ($^{134}$La) and open triangles($^{197}$Au);
Reference  \cite{Wada97} - Black triangles;
References \cite{Kwiatkowsk98, Morley96} - light-gray filled circles (Ag);
   light-gray filled triangles ($^{197}$Au);
References \cite{Cibor00,Hagel00} - black filled circles;
Reference \cite{Ruangma} - Light- gray filled squares.
 }
\end{figure}

   As noted above, for nuclei with A$>60$, the flattening of the caloric curve  sets in at similar
   excitation energies. In Fig.~\ref{fg:kvExA}, we use all data for
   A$>60$ to present another view of the evolution of the temperature-excitation
   energy correlations.  Fig.~\ref{fg:kvExA} depicts the variation of the apparent inverse
   level density parameter with excitation energy, calculated assuming Fermi gas
   behavior,  K= $T^{2}/ (E^*/A)$. At low excitation energy the apparent inverse level
   density parameter increases from  K=8 to higher values as  predicted
   in models which take into account the change in effective nucleon mass
   \cite{Hasse,Shlomo90, Shlomo91}.
   The solid line with solid points presented on the figure shows the results of
   one such calculation \cite{Shlomo90}.

    At higher excitation, however, there is a systematic change observed. At
   excitation energies in the 3-5 MeV/nucleon range the derived values of K start
   decreasing. Although there is some scatter in the experimental points, the
   overall  dependence of  K on excitation energy  manifests the limiting
   temperature behavior seen  in Fig.~\ref{fg:CalorvMass} and demonstrates quite clearly a
   qualitative change in the excited nuclei being investigated. With increasing
   excitation the values become progressively smaller. At the highest excitation
   energies they have fallen well below the value of 8 initially derived at low
   excitation.

   For the highest excitation energy data of reference \cite{Ruangma}, the higher reported
   temperatures lead to significantly higher apparent K values. This is a
   potentially interesting behavior but is different than that derived from the
   other measurements which sample that excitation energy range.
   In reference \cite{Ruangma},
   the reported uncertainties for those points are larger than the nominal 10\%
   systematic value  which we have assumed for all points.
   Also, the effect of energy cut and application of the Tsang systematics
   \cite{Tsang} in an excitation energy  range well above that for which that systematics
  was established  may be   affecting these results.

\section{Discussion \& Interpretation}

\subsection{Limiting temperatures and Coulomb Instabilities}

\begin{figure}
\epsfig{file=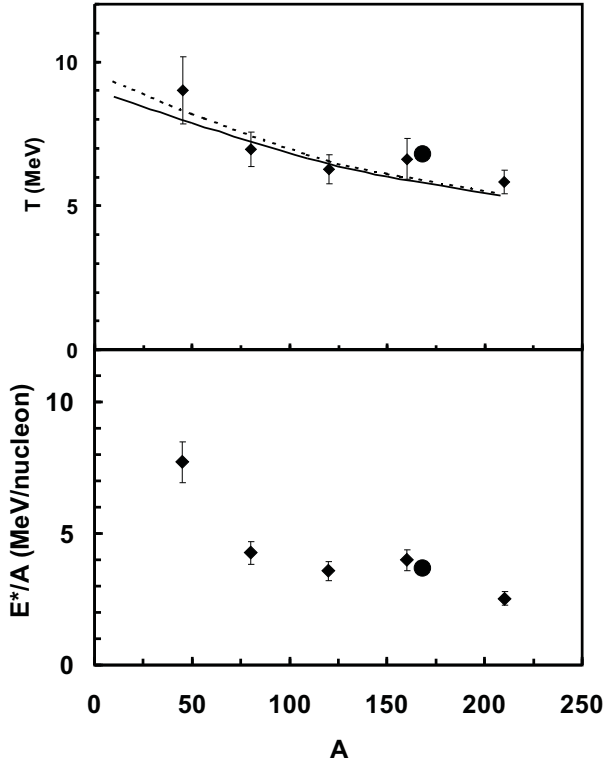,width=9.4cm,angle=0}
\caption{\label{fg:TlimitvA}
 Limiting  values of T (a) and $E^*/A$ at which $T_{limit}$ is reached
 (b) are indicated by solid diamonds. The critical temperature and
   excitation energy derived from the Fisher droplet model analysis of Elliott et al.
   \cite{Elliott01}  are represented
                 by  solid circles.  The lines in the top panel represent the calculated
   Coulomb instabiliy temperatures from references \cite{Zhang96}  (dashed line)
 and  \cite{Zhang99}   (solid line).}
\end{figure}

   As seen in the preceding section, while the curves for each mass region rise
   then flatten, the values of excitation energy and temperature at which this
   transition takes place appear to decrease with increasing mass. To further
   quantify this observation  we have applied, for each mass region, a fit  to the
   lower energy data  to determine the point of  transition from Fermi-gas-like
   behavior to the plateau region. For this purpose we used $T= \sqrt{ K (E^*/A)}$.
   Recognizing the increase in K which occurs in that region,  we have restricted
   the fits to the points near the transition. The results of these fits are shown
   in Fig.~\ref{fg:CalorvMass} (a)-(e). For the different mass windows,  the derived limiting
   temperatures are plotted in Fig.~\ref{fg:TlimitvA}(a)  and the excitation energies at
   which these limits are reached are plotted in  Fig.~\ref{fg:TlimitvA}(b).  The
   values of both T and $E^*/A$ at the transition point drop  significantly as the
   mass increases. For comparison with the derived limiting temperatures we also
   present Coulomb instability limiting temperatures calculated by Zhang et al.
   \cite{Zhang99}, employing both a
   relativistic chiral symmetry model \cite{Furnstah} and the Gogny GD1
   interaction  \cite{Zhang96}. For both,
   the calculated temperatures are in very close
   agreement with the average temperature values derived from the plateau regions
   of  the caloric curves.

   The good
   agreement between the experimental points and the values calculated using either
   the chiral symmetry model  (designated FST model in
   reference  \cite{Zhang99}) or the Gogny
   interaction favors a soft equation of state. The nuclear matter
   incompressibility in the FST model with the T1 parameterization is 194 MeV. For
   the Gogny GD1 interaction it is 228 MeV. Thus the experimental results of this
   analysis are in accord  with the incompressibilities derived from Giant Monopole
   Resonance data  \cite{Youngblood}. For finite symmetric nuclear matter the temperature
   dependence of the surface energy is taken to be that suggested by Goodman et al.
   \cite{Goodman}. The critical temperature of the FST  model with the T1 parameter set is
   14.8 MeV.  It is 15.9 MeV when the Gogny GD1 force is employed.

 \subsection{Critical Points in Nuclei}
   The question of the significance of the transition points presented in Fig.~\ref{fg:kvExA}
   may additionally be addressed using recently reported results of a Fisher
   Droplet Model analysis. The use of the Fisher droplet analysis to isolate
   possible critical behavior in nuclei has been extensively explored by the EOS
   collaboration  \cite{Elliott} and  critical parameters of the model have been extracted.
   Recently,  Elliott {\it et al.} \cite{Elliott01}
    have carried out a Fisher Droplet Model  analysis
   of the high-statistics multifragmentation data of the ISiS collaboration for 8
   GeV/c pions on $^{197}$Au  \cite{Ruangma} and shown that an impressive universal scaling of the
   data is achieved up to an excitation energy of 3.8 MeV/nucleon at which point
   the scaling is lost. This scaling behavior is found to be identical to that
   observed in three dimensional Ising model calculations which model liquid-vapor
   co-existence
     \cite{Elliott01,Mader}.
     The scaled data are interpreted as defining the
   liquid-gas co-existence line and the excitation energy of 3.8 MeV is identified
   as the critical energy for the system under investigation.

   By assuming Fermi gas behavior up to the critical point, and an inverse level
   density parameter of K=13, Elliott {\it et al.}, \cite{Elliott01} conclude that the critical
   temperature is 6.7 MeV. (In reference \cite{Ruangma} the corrected double isotope ratio
   temperatures at 3.8 MeV/nucleon excitation are$~$6.5 MeV.) In the experiment the
   excitation energy of 3.8 MeV/nucleon is associated with masses near A$\sim$168
   (see Fig.~\ref{fg:CalorvMass} (d)). Inspection of the  $\pi +~^{197}Au$ caloric curve in Fig.~\ref{fg:SumCalor} shows that
   $E^*/A$ = 3.8 MeV and T= 6.7 MeV is essentially  the point at which that caloric
   curve departs from Fermi gas-like behavior. {\it Thus this point of flattening and
   rapid departure of the caloric curve from the Fermi-gas like behavior is the
   point identified as the critical point by the droplet analysis.}

   Clearly, it would be very interesting to have data of sufficient statistics to
   carry out Fisher droplet model analyses in the different mass regions for which
   caloric curves have been  determined.  However, given that the critical point
   identified by the droplet analysis is the point of the observed departure from
   the Fermi gas behavior and flattening of the caloric curve,
   the equivalent points in other mass regions may define the critical
   energies and temperatures for those mass regions.

   The critical excitation energy and temperature determined from the droplet
   analysis of the ISiS data are  plotted in Fig.~\ref{fg:TlimitvA}. There it can be seen that the
   critical point determined in the droplet analysis is slightly higher than the transition
   point derived from the ensemble of caloric curve data in the 140-180 mass
   region. This is a direct reflection of the fact that the deviation of the ISiS
   results from the Fermi gas-like behavior occurs at higher excitation and
   temperature than indicated by the totality of experiments providing information
   in this mass region,  see Fig.~\ref{fg:CalorvMass}(d). Similarly, our recently published results
   for collisions at 47Mev/A projectile energy
    \cite{Cibor00,Hagel00}
   indicate, for the A=100-140
   region, a slightly higher critical excitation energy and critical temperature
   than is obtained from  the totality of the data in that mass region. Such
   comparisons with selected experiments emphasize the importance of accurate
   determinations of both $E^*/A$ and T to the quantitative establishment of the
   critical points.

\section{Summary and Conclusions}

   Data from a number of different experimental measurements have been combined to
   construct caloric curves for five regions of nuclear mass. These curves are
   qualitatively similar and exhibit plateaus at higher excitation energies. For
   the A$\sim$160 region, the critical point identified by a recent Fisher droplet model
   analysis  \cite{Elliott01} is observed to coincide  with the point of the observed departure
   from Fermi gas-like  behavior  and flattening in the caloric curve. This
   information is used to derive possible critical points from the caloric curves
   for other mass regions. These   temperatures and excitation energies are seen to
   decrease with increasing nuclear mass. The values are in very good agreement
   with results of recent calculations employing either a relativistic mean field
   treatment or the Gogny interaction. This agreement favors a soft equation of
   state with an incompressibility of 194-228 MeV and a critical temperature of
   14.8-15.9 MeV for symmetric nuclear matter. It should be noted, however, that
   the calculations are made for beta stable nuclei while the experiments tend to
   produce nuclei somewhat away from beta stability, on either side depending upon
   the system studied and the first stage reaction dynamics. The fluctuations which
   this might cause in the experimental results are comparable to the assumed
   systematic uncertainties in the measurements
   \cite{Besprovany,Chomaz}.

   The combined observations suggest near achievement of liquid-gas equilibrium
   analogous to that assumed in the Coulomb instability calculations. Reaching such
   a condition in these rapidly evolving systems may require that the collision
   dynamics leads to a rapid filling of the available phase space-as suggested in
   recent discussions of apparent chemical equilibrium in experiments looking for
   the ``other" phase change at relativistic energies \cite{Heinz00}.

   As indicated, it would be very interesting to have data of sufficient statistics
   to carry out Fisher droplet model analyses in the different mass regions for
   which caloric curves have been  determined. Still it is worth noting that, while
   Fisher Droplet analyses may provide the essential demonstration of critical
   behavior, a precise identification of the excitation energy and temperature at
   the critical point will continue to rely on measurements of the type surveyed
   here.

   Extension of such measurements to nuclei with very different N/Z ratios would
   also be very interesting. Significant differences in limiting temperatures
   should be seen in more asymmetric systems and the order of the phase transition
   is expected to change
    \cite{Besprovany,Chomaz,Muller}.
    For the systems already studied, the
   differences in the entrance channel isospins and in the first stage dynamics
   lead to some variation of the isospin of the fragmenting nuclei. However the
   systematic uncertainties in the present measurements are such that sensitivity
   to this variable is not obvious. With radioactive beams it should be possible to
   employ caloric curve measurements to determine the critical parameters for quite
   asymmetric nuclei, thus testing the isospin dependence of the equation of state.
   For such beams, the intensity limitations far from stability will mean that
   caloric curve measurements will be inherently easier to obtain than will the
   high statistics data needed for a droplet analysis. Caloric curve measurements
   should continue to be an important tool for probing the equation of state.

   Note added: Following submission of this paper, Srivastava {\it et al.}  \cite{Srivastava}
   have submitted a preprint
    reporting
   results   of a systematic analysis of the moments of the fragment size
   distributions in their
   EOS data. This analysis also indicates a decrease in temperature and excitation
   energy with increasing mass, which they attribute to Coulomb effects.   Also,
   Dorso and Bonasera \cite{Dorso01a} have recently published results of an analysis of
   molecular dynamics calculations which identified the region of entry into the plateau as the
   region where fluctuations are maximal and critical behavior could be expected.

\section{Acknowledgements}
   We thank W. Trautmann, D. Cussol, A. Ruangma, A. Hauger and B. Srivastava for
   providing us with numerical values of their data.  We appreciate very useful
   conversations with Shalom Shlomo and H. Q. Song. This work was supported by the
   U S Department of energy under Grant DE-FG03-93ER40773 and by the Robert A.
   Welch Foundation.

\begin{table*}
\caption{Summary of   Measurements Included in Analysis}
\begin{center}
\begin{tabular}{lllllll}
\multicolumn{2}{c}{\bf Reference} &{\bf Reactions} & \multicolumn{2}{c}{\bf Temperature} &
\multicolumn{2}{c}{\bf Excitation Energy} \\
       &                &     	  & {\bf Method}& {\bf Correction}& {\bf Method}& {\bf Correction} \\		
6    & Hagel {\it et al.}&   19,35 MeV/A N+Sm& He Slope & Cascade    & Momentum  & None \\	
      &                    &                                  &                 & Correction &  Transfer & \\ \hline 				
7    & Wada {\it et al.}&  30Mev/A O,S+Ag & He Slope & Cascade 	 & Momentum & None \\	
      &                    &                                  &                 & Correction &  Transfer & \\ \hline 				
 8   &Cussol {\it et al.}& 36-65Mev/A Ar+Al & He Slope &	Cascade & Momentum  & None \\	
      &                    &                                  &                 & Correction &  Transfer & \\ \hline 				
9 & Chulick {\it et al.}&	 10Mev/A C+Sn	    & He Slope &	None        & Momentum  & None \\
      &                    &                                  &                 &                    &  Transfer & \\ \hline 				
														
10 & Gonin {\it et al.}& 11Mev/A Ni+Mo	    & H,He Slope &	 Subtraction & Calorimetry & None \\	 \hline
														
11  &	Pochadzalla & 1000Mev/A Au+Au & HeLi Isotope &	QSM Model & Calorimetry & SMM  \\ 	
12  & 	{\it et al.} Odeh                     & & Ratio & & & Backtrace \\ \hline												
13  &	Hauger {\it et al.}& 1A GeV Kr, La, Au+C & HHe Isotope  & SMM Model	 & Calorimetry & Pre-Eq.  \\
       &                          &                                      &  Ratio & & & Removed \\  \hline
14  & Wada {\it et al.}   &  35Mev/A Cu+Au	 & HHe Isotope  & QSM Model	 & Calorimetry & Pre-Eq.  \\
       &                          &                                      &  Ratio & & & Removed \\  \hline
15,16  & Kwiatkowski &	 4.8 GeV He+Ag,Au & HHe Isotope  & SMM E Window & Calorimetry & Pre-Eq.  \\	
  &  {\it et al.}           &                                    &   Ratio                        & Correction &  & Removed \\ \hline				
17,18  & Cibor {\it et al.}  & 47Mev/A C, Ne, Ar, Zn & HHe Isotope  & None	 & Calorimetry & Pre-Eq.  \\
   & Hagel {\it et al.}  & + Med. Mass and Au          & Ratio & & &  Removed \\	 \hline									
19  & Ruangma & 8 GeV/c pion+Au & HHe Isotope  & Tsang	& Calorimetry & Pre-Eq.  \\ 	
      &  {\it et al.}        &                                &   Ratio                     & Systematics	 &  & Removed \\ 	 \hline				
\end{tabular}
\label{tb:DataRef}
\end{center}
\end{table*}

\end{document}